\documentclass{emulateapj}

\usepackage{epsfig}

\newcommand{\etal}{\mbox{et al.}}
\newcommand{\ergcms}{erg cm$^{-2}$ s$^{-1}$}
\newcommand{\ergs}{erg s$^{-1}$}
\newcommand{\phcms}{ph cm$^{-2}$ s$^{-1}$}

\newcommand{\msun}{$M_{\odot}$}

\newcommand{\chandra}{{\it Chandra}}
\newcommand{\rosat}{{\it ROSAT}}

\newcommand{\sgrastar}{\mbox{Sgr A$^*$}}

\newcommand{\newbe}{\mbox{CXOGC J174516.1-290315}}
\newcommand{\alvin}{\mbox{X174516.1}}

\shortauthors{Muno \etal}
\shorttitle{Isolated Galactic Center Supergiants}

\begin{document}

\title{Isolated, Massive Supergiants near the Galactic Center}
\author{M. P. Muno,\altaffilmark{1,2} 
G. C. Bower,\altaffilmark{3}
A. J. Burgasser,\altaffilmark{4,5,6}
F. K. Baganoff,\altaffilmark{7} 
%G. P. Garmire,\altaffilmark{6}
M. R. Morris,\altaffilmark{1} and
%G. R. Ricker,\altaffilmark{5}
W. N. Brandt\altaffilmark{8}
}

\altaffiltext{1}{Department of Physics and Astronomy, University of California,
Los Angeles, CA 90095; mmuno@astro.ucla.edu}
\altaffiltext{2}{Hubble Fellow}
\altaffiltext{3}{Radio Astronomy Lab, University of California, Berkeley,
CA 94720}
\altaffiltext{4}{Department of Astrophysics, Division of Physical
Sciences, American Museum of Natural History, Central Park West at
79$^{th}$ Street, New York, NY 10024}
\altaffiltext{5}{Spitzer Fellow}
\altaffiltext{6}{Visiting Astronomer at the 
Infrared Telescope Facility, which is
operated by the University of Hawaii under Cooperative Agreement
NCC 5-538 with the National Aeronautics and Space Administration,
Office of Space Science, Planetary Astronomy Program} 
\altaffiltext{7}{Kavli Institute for Astrophysics and Space Research,
Massachusetts Institute of Technology, Cambridge, MA 02139}
\altaffiltext{8}{Department of Astronomy and Astrophysics, 
The Pennsylvania State University, University Park, PA 16802}

\begin{abstract}
We have carried out a pilot project to assess the feasibility of using
radio, infrared, and X-ray emission to identify young, massive stars 
located between 1 and 25 pc from the Galactic center. 
We first compared catalogs compiled from the Very Large Array, the 
{\it Chandra X-ray Observatory}, and 2MASS. We identified two
massive, young stars: the previously-identified star that is associated with
the radio HII region H2, and a newly-identified star that we refer to 
as \newbe. The infrared spectra of both stars exhibit
very strong Brackett-$\gamma$ and He I lines, and resemble those of massive
supergiants that have evolved off of the main sequence, but not yet 
reached the Wolf-Rayet phase. We estimate that each star has a bolometric 
luminosity $\ga$$10^{6}$$L_{\odot}$. These two stars are also associated with
bright mid-infrared sources from the {\it Midcourse Space Experiment}
survey, although the origin of this emission is uncertain. Likewise, the 
detection of these two sources in X-rays is surprising, because stars at 
similar evolutionary
states are not uniformly bright X-ray sources. Therefore, we suggest that 
both stars are in binary systems that contain either OB stars whose
winds collide with those of the luminous supergiants, or compact objects
that are accreting from the winds of the supergiants.
We also identify X-ray emission from a nitrogen-type Wolf-Rayet star
and place upper limits on the X-ray luminosities of 
three more evolved, massive stars that previously have been identified 
between 1 and 25 pc from Sgr A*. Finally, we briefly discuss the implications 
that future searches for young stars will have for our understanding 
of the recent history of star formation near the Galactic center.
\end{abstract}

\section{Introduction}

In contrast to the Galactic Bulge, the inner 300 pc of the Galaxy
is experiencing ongoing star formation. This is dramatically displayed 
by the $\ga$60 ultra-compact HII regions in the
giant molecular cloud Sgr B2 \citep{dgg98}, 
and by three young, dense clusters of massive stars 
\citep[the Arches, the Quintuplet, and the Central Parsec; e.g.,][]{kra95,fig99}.
The average star formation rate is estimated to be $\sim$0.02 \msun\ 
yr$^{-1}$, or $\sim$1\% of the total Galactic value \citep{fig04}. About
half of the recent star formation can be accounted for by the 
three massive star clusters mentioned above, but this implies that 
the sites of formation for the other half of the young stars in the 
central 300 pc has not yet been identified. One possibility is that 
star clusters less massive than the Arches and Quintuplet 
($\sim$$10^{4}$ \msun) have already begun to dissipate, and are difficult 
to identify against the dense background of older stars near the Galactic 
center \citep{pz02}. 
Alternatively, a half-dozen relatively isolated HII regions near the 
Galactic center have been associated with individual young, massive 
stars \citep{cot99}, 
which could suggest that stars can also be formed in small associations.
Our understanding of how stars form near the Galactic 
center would be greatly helped by identifying more of the young stars 
expected to exist there.

We have carried out a radio survey at 8.4 GHz with Very Large Array (VLA)
of the region between 3\arcmin\ and 10\arcmin\ from the super-massive
black hole \sgrastar, in order to identify counterparts to the 
$\approx$2000 X-ray sources identified using the {\it Chandra X-ray 
Observatory} \citep{mun03a}. The general properties of the radio sources
detected in this survey will be described elsewhere. Here, we report 
the properties of two radio and X-ray sources that also had counterparts 
among the $\approx$18,000 stars in the same region from the Two Micron All-Sky 
Survey (2MASS; \S2.1). We identify one, \newbe, as a new emission-line 
star located 10 pc in projection from the Galactic center.
The second is a previously-identified luminous, young star associated with
radio HII region H2 \citep{yzm87,zhao93,fig95,cot99}. 

We also present 
an assessment of whether X-ray and radio surveys could be generally 
effective for identifying luminous young stars (\S2.2). We find X-ray emission
from two of five massive, emission-line stars that were previously identified 
through infrared spectroscopy of stars located near HII regions by 
\citet{cot99} (H2 and a WN6 star associated with SgrA-A), and provide 
updated positions for all five stars based on the 2MASS catalog. We 
discuss the origin of the X-ray emission from the young stars in our sample,
and the implications that the detection of the new young, massive star has
for understanding the star formation history near the Galactic center (\S3).

\begin{figure}
%\epsscale{1.0}
%\plotone{f1.eps}
\centerline{\epsfig{file=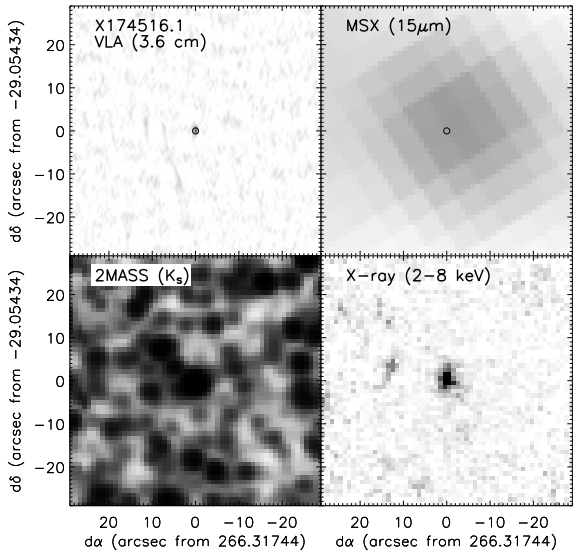,width=\linewidth}}
\caption{
Images of \newbe\ from the VLA at 3.6 cm, from the MSX survey at 
15 $\mu$m, from the 2MASS survey at 2.2 $\mu$m ($K_s$), and 
from \chandra\ integrated over the 2--8 keV bandpass. The star is 
indicated with the $+$, and is detected as a point source in all of 
the images. The radio image is displayed on a linear scale with 
minimum intensity of 0 mJy per beam and maximum of 3 mJy per beam. The 
MSX and \chandra\ images are displayed using logarithmic scales. 
The intensity at 15 $\mu$m is 33 Jy, and the 2--8 keV flux is
$5.6\times10^{-15}$~\ergcms. The 
2MASS image is displayed using equalized color histograms.
The near-infrared sources has a magnitude of 
$K_s$=7.9. 
}
\label{fig:newimg}
\end{figure}

\begin{figure}
%\epsscale{1.0}
%\plotone{f1.eps}
\centerline{\epsfig{file=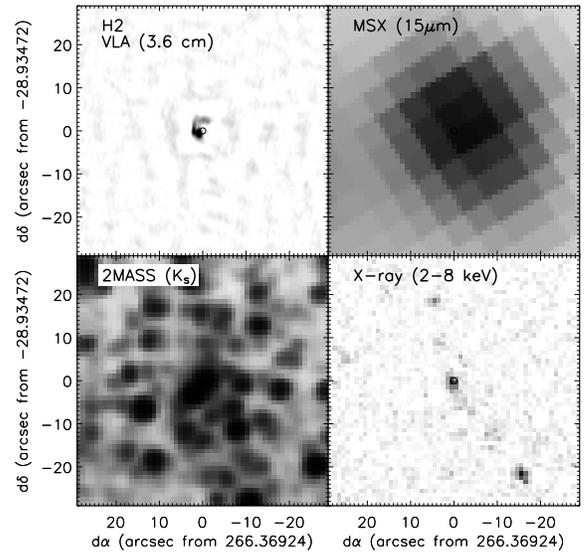,width=\linewidth}}
\caption{
Same as Figure~1, for H2. The radio image is displayed on a linear scale with 
minimum intensity of 0 mJy per beam and maximum of 50 mJy per beam. The 
remaining images are displayed using the same scalings as Figure~1. The radio emission is clearly extended, 
but it appears as a point source in the mid-infrared,
infrared, and X-ray. The intensity at
15 $\mu$m is 220 Jy. The near-infrared sources has a magnitude of 
$K_s$=9.2. The X-ray source has a 2--8 keV flux of $1.7\times10^{-15}$~\ergcms.
}
\label{fig:h2img}
\end{figure}

\section{Observations}

\subsection{Radio and X-ray Selected Young Stars}

We identified two young stars by comparing a catalog of discrete 
sources compiled from Very Large Array (VLA) observations at 8.4 GHz 
(3.6 cm; G. C. Bower \etal, in prep) to the \chandra\ and the 2MASS catalogs. 
The VLA observations were taken in the B-array during 2002 August. 
They were sensitive to sources as faint as 250 $\mu$Jy that were 
located between 5\arcmin\ and 10\arcmin\ of 
\sgrastar, although the sensitivity declined to $\approx$3 mJy within a 
few arcminutes of \sgrastar\ because of the bright diffuse radio emission
there. The VLA observations identified 30 unresolved sources, and 
55 extended sources. The positions of the unresolved sources were accurate
to 0\farcs1.

We searched for X-ray counterparts using the catalog in \citet{mun03a}.
The uncertainties in the positions of the X-ray sources in the
\chandra\ catalog are accurate to 0\farcs3 at the aim point of the 
observation (i.e., \sgrastar), but within 5--10\arcmin\ from 
\sgrastar\ the average uncertainty was $\sim$1\arcsec. Since most of the 
radio sources were found at larger offsets, we searched for X-ray and 
radio sources that were located within 1\arcsec\ in projection. 
We found two matches between the unresolved radio sources and the X-ray 
catalog within a radius of 1\arcsec.
By adding random shifts to the relative alignments of the two catalogs, 
we determined that 1 of these matches was likely to be spurious. 
Within the extents of the diffuse 
VLA sources, we find 5 X-ray counterparts, although three or four should be 
random associations. 
%Within 0\farcs5: 1 diffuse, 0.75 expected; 2 point, < 0.2 expected 
%Within 1\arcsec: 2 diffuse, 1.8 expected; same
%Within 2\arcsec: 5 diffuse, 4.5 expected; 4 point, < 0.2 expected
%point sources are possibly significant to 6 arcsec or so.

We then compared the locations of the 7 X-ray sources with the 2MASS
catalog, and found that 2 had infrared counterparts within 1\arcsec\
of the X-ray sources: 
CXOGC J174528.6--285605, which is coincident with a previously-identified 
young star in the bright ($\approx$450 mJy at 8.4 GHz), extended radio 
nebula H2 \citep{yzm87,zhao93,fig95,cot99}, 
and the new source \newbe\ (hereafter referred to as \alvin), which is  
associated with a 2.3$\pm$0.3 mJy unresolved radio source. The chance that
one of these radio and X-ray sources randomly would have a 2MASS counterpart 
within 1\arcsec\ is $<$15\%; if we restrict ourselves to 2MASS sources 
with $K_s$$<$10, the chance is $<$4\%. The locations and $JHK_s$ magnitudes 
of the 2MASS counterparts are listed in Table~\ref{tab:prop}, and radio,
infrared, and X-ray images of the sources are displayed in 
Figures~\ref{fig:newimg} and \ref{fig:h2img}.

\begin{deluxetable*}{lcccccccc}
\tablecolumns{9}
\tablewidth{0pc}
\tabletypesize{\scriptsize}
\tablecaption{Basic Properties of Young Stars Near the Galactic Center\label{tab:prop}}
\tablehead{
\colhead{Name} & \colhead{ra} & \colhead{dec} & \colhead{Offset} & 
\colhead{$J$} &
\colhead{$H$} & \colhead{$K_s$} & \colhead{Net} & \colhead{Bkgd.} \\
\colhead{} & \multicolumn{2}{c}{(J2000)} & \colhead{(arcmin)} & \colhead{} &
\colhead{} & \colhead{} & \multicolumn{2}{c}{X-ray Counts}
}
\startdata
\alvin & 266.31744 & --29.05434 & 5.9 & 11.49$\pm$0.02 & 9.12$\pm$0.03 & 7.89$\pm$0.02  & 706$\pm$31 & 170$\pm$ 4 \\
H2 & 266.36924 & --28.93472 & 5.0 & 14.3 & 11.26$\pm$0.03 & 9.22$\pm$0.03 &  201$\pm$18 &  86$\pm$ 2 \\
H8 & 266.40761 & --28.95448 & 3.2 & 15.6$\pm$0.1 & 12.30$\pm$0.08 & 10.30$\pm$0.05 &  $<$17 &  52$\pm$ 1 \\
H5 & 266.41381 & --28.88923 & 7.1 & 14.73$\pm$0.04 & 11.68$\pm$0.04 & 10.19$\pm$0.05 & $<$27 & 257$\pm$ 7 \\
SgrA-A & 266.46089 & --28.98879 & 2.6 & 15.59$\pm$0.07 & 12.45$\pm$0.02 & 10.90$\pm$0.03 &  52$\pm$12 &  80$\pm$ 2  \\
SgrA-D & 266.46477 & --29.00643 & 2.5 & 17.6 & 15.1 & 11.20$\pm$0.04 & $<$19 &  53$\pm$ 1
\enddata
\tablecomments{Positions are accurate to 0\farcs3. The offset is relative
to \sgrastar. Uncertainties on the 
infrared magnitudes are provided where they are available from the 2MASS 
catalog.}
\end{deluxetable*}

\begin{deluxetable}{lccc}
\tabletypesize{\scriptsize}
\tablecolumns{4}
\tablewidth{0pc}
\tablecaption{Emission Lines from Transitional Stars\label{tab:lines}}
\tablehead{
\colhead{Transition} & \colhead{Centroid} & \multicolumn{2}{c}{EW (\AA)} \\
\colhead{} & \colhead{($\mu$m)} & \colhead{\alvin} & \colhead{H2}
}
\startdata
He I $2s^1S-2p^1P^o$ & 2.058 & 33$\pm$1 & 5.0$\pm$0.3 \\ [5pt]
He I $2p^3P^o-4s^3S$ & & & \\
+ He I $3p^1P^o-4s^1S$ & 2.115 & 1.8$\pm$0.6 & 6.9$\pm$0.6 \\
+ N III/ C III 8--7 & & & \\ [5pt]
Mg II $5s^2S_{1/2}-5p^2P^o_{1/2}$ & 2.142 & 2.0$\pm$0.7 & $<$1.9 \\ [5pt]
H I 7--4 & & & \\
+ He I $7f^3F^o-4d^3D$ & 2.167 & 25$\pm$1 & 22.5$\pm$0.3 \\
+ He I $7f^1F^o-4d^1D$ & & & \\
%He II 10--7 & 2.1890^\tablenotemark{a} & 0.5 \\
%+ Fe II $f^4G_{5/2}-4p^4G_{\5/2}$ & & \\
\enddata
%\tablenotetext{a}{Centroid fixed.}
\tablecomments{The spectra were taken with $R$=150, so at 2 $\mu$m the
resolution is $\approx$0.01 $\mu$m.}
\end{deluxetable}

\begin{figure}
%\epsscale{1.0}
%\plotone{spex_spec.ps}
\centerline{\epsfig{file=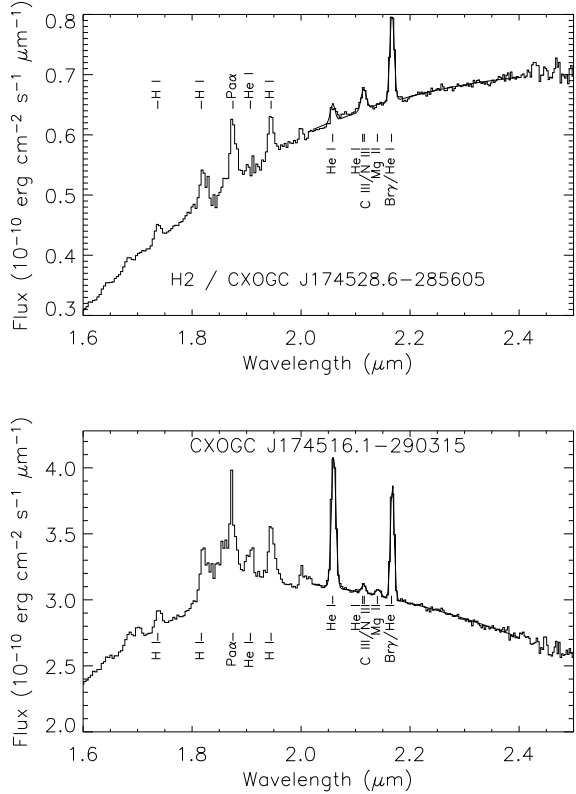,width=\linewidth}}
\caption{Infrared spectrum of the new emission-line star, taken with
SpeX on IRTF. The spectrum we obtained extends to shorter 
wavelengths, but the strong absorption toward the Galactic 
center cut-off the spectrum short-ward of 1.7$\mu$m. 
The strong H, He, and Mg lines
are reminiscent of Of and LBV stars in \citet{mor96}.}
\label{fig:spex}
\end{figure}

To determine the nature of the infrared counterparts, we obtained
low-resolution ($R$=150) infrared spectra with SpeX on the Infrared Telescope 
Facility \citep[IRTF; Figure~\ref{fig:spex};][]{ray03}. The spectra
were taken on 2004 September 6 (UT). The weather was clear, with 
0\farcs7 seeing at $J$. The acquisition and reduction of the spectra 
was carried out as described in \citet{bur04}. To account for background 
emission, a spectrum was extracted from an adjacent region along the slit
that did not contain obvious stellar sources, and then subtracted from 
the source spectrum.
The flux values for the spectra were scaled to the 2MASS magnitudes 
using the filter and optical responses and the atmospheric
transmission in \citet[][]{cut03}.

Both stars exhibit prominent emission lines (Figure~\ref{fig:spex}). 
The strong line emission between 1.8 and 2.0 $\mu$m cannot be resolved
into distinct lines in our low-resolution spectrum. The strongest lines
are probably from the Brackett series of H \citep{fig99}, but their 
equivalent widths cannot be measured because it is not possible to 
determine the shape of the underlying continuum. We can resolve lines from 
the continuum at wavelengths
longer than 2.0 $\mu$m. In Table~\ref{tab:lines}, we tentatively identify 
strong lines from He I and H Brackett-$\gamma$, and weaker lines that 
could be from CIII, NIII, and/or Mg II \citep[see][]{mor96,fig99}. 
We measured the equivalent widths of the lines by modeling the continuum
between 2.1 and 2.4 $\mu$m as a second order polynomial, and the 
lines as Gaussians. The equivalent widths also are listed in 
Table~\ref{tab:lines}.
The presence of strong H and He emission lines indicates that both sources 
are hot, massive stars.

The radio continuum and infrared line emission suggested that there could 
be significant amounts of warm gas and dust surrounding these young stars
\citep[e.g.,][]{cla03}.
We searched for this in the mid-infrared, using version 2.3 of
The Midcourse Space Experiment (MSX) Point Source Catalog \citep{egan03}.
\alvin\ and H2 are the only two sources detected in the X-ray and 
2MASS catalogs that also has a counterpart within 1\farcs5 in the MSX 
catalog. We find that there is a $<$15\% chance of finding one random 
association between the three catalogs. 
H2 is coincident with MSX6C G259.9845+00.0275, and has fluxes of 
33, 134, 220, and 490 Jy at 8.3, 12.1, 14.7, and 21.3 $\mu$m, 
respectively. \alvin\ is coincident with 
MSX6C G359.8590+00.0036, and has fluxes of 2, 14, 33, and 102 Jy
at 8.3, 12.1, 14.7, and 21.3 $\mu$m. The fluxes are accurate to
5\%, although the angular resolution of MSX was $\approx$20\arcsec,
so the mid-infrared emission may represent a blend of nearby stars.

As we describe in \S3, the strength of the X-ray emission from these 
young stars is a bit surprising. Therefore, in order to determine its origin, 
we determined the X-ray intensities, spectra, and light curves for each star 
using the acis\_extract routine from the Tools for X-ray Analysis 
(TARA)\footnote{www.astro.psu.edu/xray/docs/TARA/} and CIAO version 3.2.
The techniques are described in \citet{mun04b}, and the dates and exposure 
times for the observations used are listed in \citet{mun05b}. In brief, 
for each source
we extracted source events from within the 90\% contour of the point spread 
function (PSF), and background events using annular regions that excluded
known point sources and discrete filamentary features. 
We list the net and background X-ray counts 
from each source in Table~\ref{tab:prop}. These values supersede those in 
\citet{mun03a}.

\begin{figure}
%\epsscale{1.0}
%\plotone{f1.eps}
\centerline{\epsfig{file=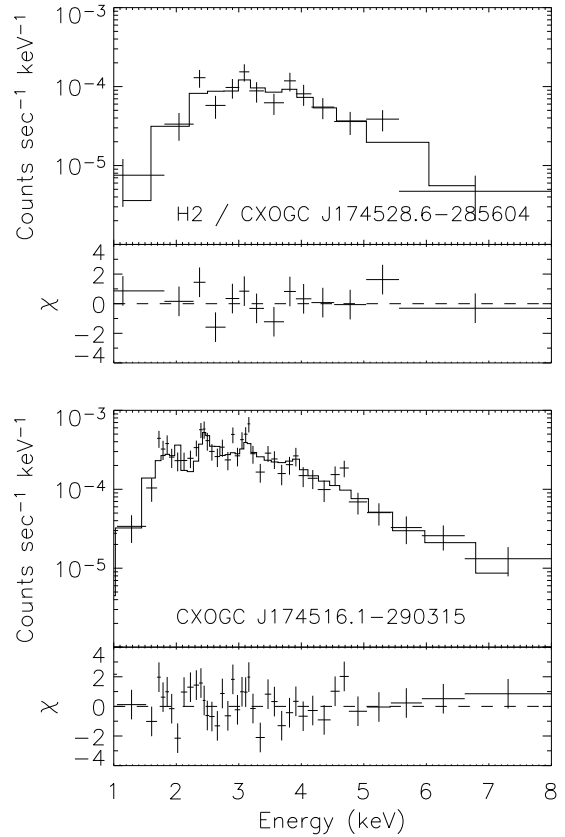,width=0.9\linewidth}}
\caption{
The X-ray spectra of the two bright X-ray sources, H2 and \alvin.
The top panels display the spectra in 
detector counts as a function of energy, so the shape of the source 
spectra are convolved with the detector response. The bottom panels contain 
the differences between the
data and the best-fit thermal plasma models, divided by the 
Poisson uncertainties on the data points.
}
\label{fig:xspec}
\end{figure}

We then produced source and 
background spectra from the event lists, computed the effective area
using the CIAO tool {\tt mkarf}, and obtained the response functions 
tabulated by \citet{tow02a}.
The mean fluxes of the sources did not change during the first 85\% the 
exposure taken through 2003 June \citep[][although see below]{mun04b}, so 
we used summed spectra for each source.
We modeled the spectra using XSPEC version 11.3.1 \citep{arn96} as 
a thermal plasma \citep{mew86} absorbed by interstellar gas and dust. 
The free parameters in this model were the column of interstellar 
gas ($N_{\rm H,ISM}$), 
the plasma temperature ($kT$) and normalization
(proportional to the emission measure $K_{\rm EM} = \int n_e n_H dV$). 
The optical depth of dust was set to 
$\tau = 0.485 \cdot N_{\rm H}/(10^{22} {\rm cm}^{-2})$, and the halo area
to 100 times that of the PSF \citep{bag03}. 
The best-fit spectral parameters and derived luminosities 
for the two detected sources are listed in Table~\ref{tab:spec}, and the 
spectra are displayed in Figure~\ref{fig:xspec}.

Finally, massive stars, particularly those in binaries, are often
variable X-ray sources. Therefore, we examined whether X-ray light curves for
H2 and \alvin\ were consistent with a constant mean flux using 
Kolgomorv-Smirnov (K-S) and $\chi^2$ tests. 
The photon flux from \alvin\ increased
by a factor of 3 from $(3.1\pm0.2)\times10^{-6}$ \phcms\ between 
1999 and 2003 to $(7.8\pm0.7)\times10^{-6}$ \phcms\ during 2004 July.
The probability that this increase in flux resulted from a constant
flux was $<2\times10^{-7}$ under a $\chi^2$ test. 
We find no evidence that the hardness of the spectrum varied during the
outburst. We defined a hardness ratio $(h-s)/(h+s)$ using $s$ as the 
number of counts in the 0.5--3.0~keV band, and $h$ as the counts in the
3.0--8.0 keV band. The average hardness from 1999 through 2003 was
0.02$\pm$0.06, while that during 2004 July was $-$0.01$\pm$0.09, so there
is no evidence for spectral variations coincident with the increase in flux.
On shorter time scales, there appears to have been a flare near or 
before the start of the observation on 2004 July 6, which decayed with a time
scale of $\approx$1 hr. During that observation, the chance that the 
data were produced by a constant flux was $<$0.3\% according to a K-S
test.

\begin{deluxetable*}{lcccccc}
\tabletypesize{\scriptsize}
\tablecolumns{7}
\tablewidth{0pc}
\tablecaption{X-ray Spectra of Young Stars near the Galactic 
Center\label{tab:spec}}
\tablehead{
\colhead{Infrared} & \colhead{$N_{\rm H}$} & 
\colhead{$kT$} & \colhead{Norm} & \colhead{$\chi^2 / \nu$} &
\colhead{$F_{\rm X}$} & \colhead{$L_{\rm X}$} \\
\colhead{ID} & \colhead{($10^{22}$ cm$^{-2}$)} &
\colhead{(keV)} & \colhead{} & \colhead{} & 
\colhead{(\ergcms)} & \colhead{(\ergs)}
}
\startdata
H2 & $7.7_{-1.4}^{+1.5}$ & $1.2_{-0.2}^{+0.2}$ & $1.0_{-0.5}^{+1.2}\times10^{-4}$ & 11/11 & $5.6\times10^{-15}$ & $1.4\times10^{33}$ \\
SgrA-A & 6.0\tablenotemark{a} & $>2$ & $4_{-2}^{+3}\times10^{-6}$ & 5/4 & $1.7\times10^{-15}$ & $5\times10^{31}$ \\
\alvin & 4.7$_{-0.3}^{+0.3}$ & 1.3$_{-0.1}^{+0.1}$ & $1.5_{-0.4}^{+0.3}\times10^{-4}$ & 50/37 & $1.7\times10^{-14}$ & $1.9\times10^{33}$
\enddata
\tablecomments{The fluxes and luminosities are quoted for the 0.5--8.0 keV
band. However, as a result of the high extinction toward the Galactic 
Center, most of the observed flux is detected between 2 and 8~keV. The 
luminosities in this band are factors of 2--4 below those in the table.}
\tablenotetext{a}{Parameter fixed because there were few counts from this 
source.}
\end{deluxetable*}

The X-ray light curve 
from H2 has a $\approx$30\% chance of being produced by a constant mean flux
under both tests. However, because the count rate from H2 is lower than
that from \alvin, we cannot exclude the hypothesis that a flux variation 
with a factor of $\la$3 amplitude occurred between 1999 and 2004.

\subsection{X-ray Counterparts to Known Emission-Line Stars}

\begin{figure}
%\epsscale{1.0}
%\plotone{f1.eps}
\centerline{\epsfig{file=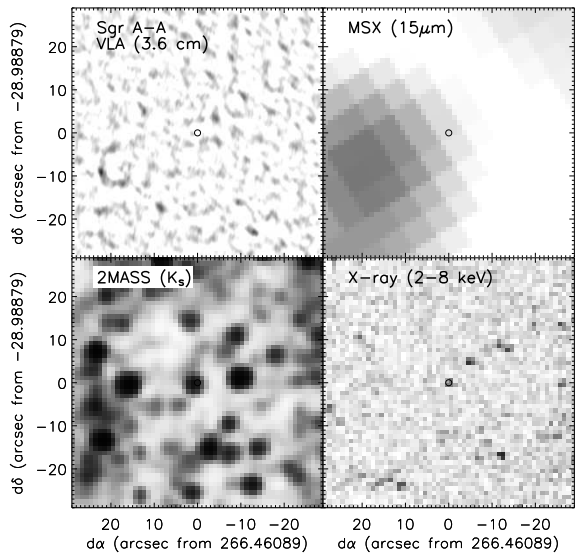,width=\linewidth}}
\caption{
Same as Figure~1, for the WR star near Sgr A-A. 
The radio image is displayed on a linear scale with 
minimum intensity of 0 mJy per beam and maximum of 50 mJy per beam. The 
remaining images are displayed using the same scalings as Figure~1. The WR 
star is visible as a point source in the infrared and X-rays. 
The upper limit to the flux density of a 
point-like radio source coincident with the WR star is $<$2.4 mJy. The HII 
region Sgr A-A can be seen in the mid-infrared, located 20\arcsec\ east 
and 10\arcsec\ south of the WR star. The VLA observations have resolved 
out the diffuse radio emission. 
}
\label{fig:sgraa_img}
\end{figure}

After realizing that H2 had been previously identified as a young
star, we returned to previous searches for such stars to determine 
if others were also strong X-ray sources. For the purposes of 
this paper, we are interested in relatively isolated stars,\footnote{The 
X-ray properties of the stars in the central parsec will be reported in 
a future publication.} so we searched for X-ray counterparts to 
the four emission-line stars (besides H2) in \citet{cot99}. We first 
compared the locations and $JHK$ magnitudes of the \citet{cot99} stars to 
the 2MASS catalog in order to refine their locations. We found that the 
published positions of the stars disagreed with their 2MASS 
counterparts by 1--3\arcsec. Therefore, we report updated positions and 
the $JHK_s$ magnitudes of the emission-line stars in Table~\ref{tab:prop}.
Using the updated positions, we found that one more of the remaining four
emission-line stars has an X-ray counterpart in our \chandra\ image
within the 0\farcs5 uncertainty of the X-ray catalog: the WN6 star 
SgrA-A.\footnote{We note that the inaccurate positions in \citet{cot99} led us 
to report initially that none had counterparts \citep{mun04b}.} 
This star was not detected in our radio and 
X-ray comparison (\S2.1), because the infrared and X-ray sources are 
significantly offset from the nearby radio HII region. We have displayed
images of SgrA-A in Figure~\ref{fig:sgraa_img}.

\begin{figure}
%\epsscale{1.0}
%\plotone{f1.eps}
\centerline{\epsfig{file=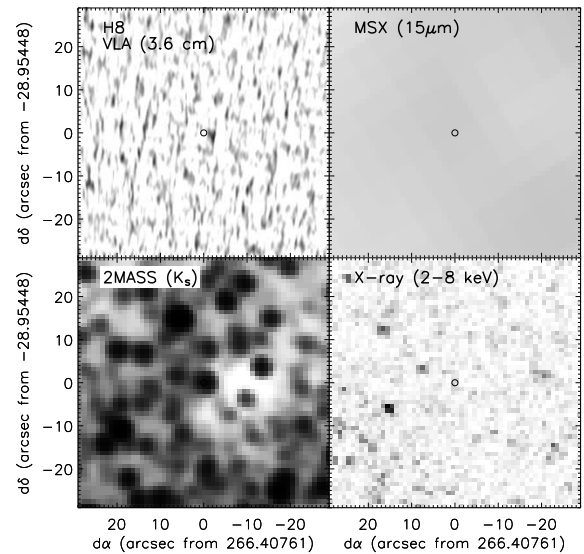,width=\linewidth}}
\caption{
Same as Figure~1, for the star associated with the radio nebula H8. 
The radio image is displayed on a linear scale with 
minimum intensity of 0 mJy per beam and maximum of 3 mJy per beam. The 
remaining images are displayed using the same scalings as Figure~1. The
star is only visible in the infrared image; it is not detected in
X-rays. The VLA observations have resolved out the emission from the 
HII region, and there upper limit to the flux density of a point-like
radio source is $<$1.2 mJy. The mid-infrared emission exhibits a diffuse 
background that is not resolved into discrete features. 
}
\label{fig:h8_img}
\end{figure}

\begin{figure}
%\epsscale{1.0}
%\plotone{f1.eps}
\centerline{\epsfig{file=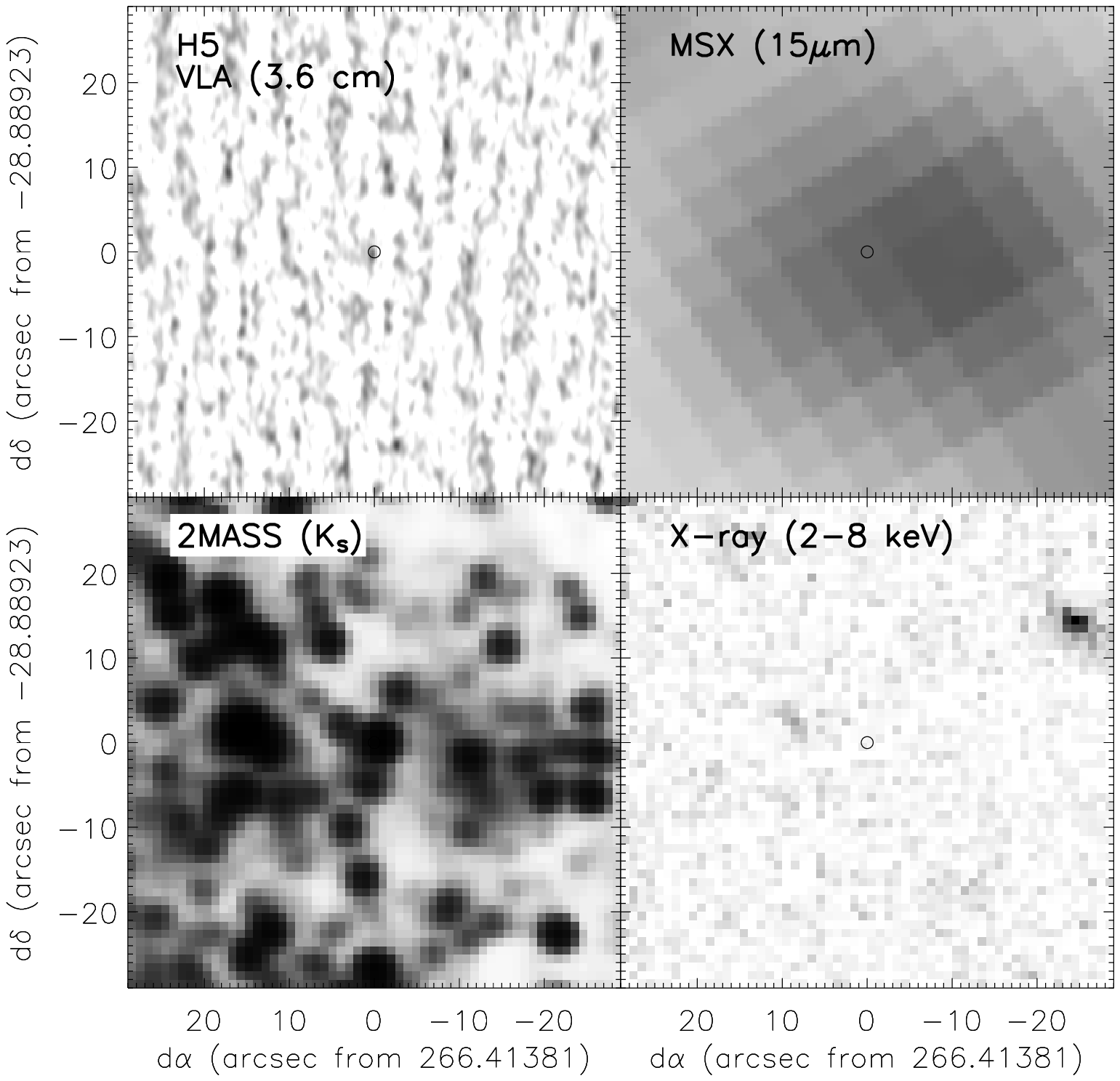,width=\linewidth}}
\caption{
Same as Figure~1, for the star associated with the radio nebula H5. 
The radio image is displayed on a linear scale with 
minimum intensity of 0 mJy per beam and maximum of 3 mJy per beam. The 
remaining images are displayed using the same scalings as Figure~1. The
star is only visible in the infrared image; it is not detected in
X-rays. The VLA observations have resolved out the emission from the 
HII region, and there upper limit to the flux density of a point-like
radio source is $<$1.6 mJy There is extended mid-infrared emission 
from the nebula.
}
\label{fig:h5_img}
\end{figure}

\begin{figure}
%\epsscale{1.0}
%\plotone{f1.eps}
\centerline{\epsfig{file=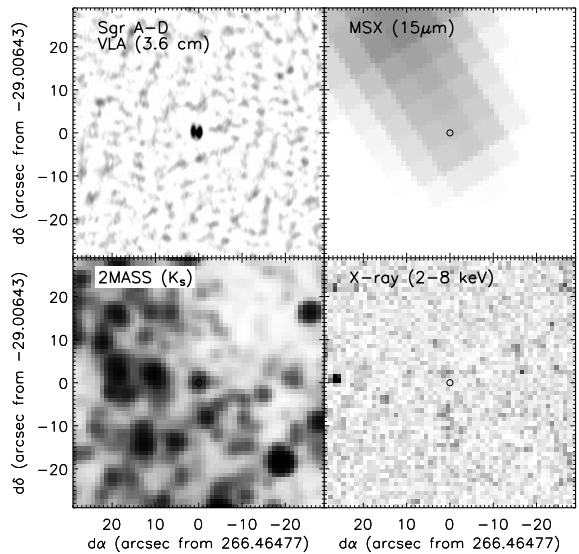,width=\linewidth}}
\caption{
Same as Figure~1, for the star associated with the radio nebula SgrA-D. 
The radio image is displayed on a linear scale with 
minimum intensity of 0 mJy per beam and maximum of 50 mJy per beam. The 
remaining images are displayed using the same scalings as Figure~1. The star is visible in the infrared image, and is
coincident with two point-like radio features with fluxes of 22 mJy.
Most of the radio emission from the HII region is resolved out, but it 
can be seen in the mid-infrared image. The star is not detected in X-rays. 
}
\label{fig:sgrad_img}
\end{figure}

We analyzed the X-ray emission from each of the four stars in the same manner 
as in \S2.2, and list their properties in Table~\ref{tab:spec}.
For the three non-detections from \citet{cot99}, we computed limits
on their count rates by extracting events from the 2MASS positions 
(Table~\ref{tab:prop}). The limits on their luminosities for a 1~keV thermal 
plasma spectrum are $1\times10^{32}$~\ergs, and for a 0.5 keV plasma are 
$2\times 10^{33}$~\ergs. Images of the stars not detected in X-rays
are displayed in Figures~\ref{fig:h8_img}--\ref{fig:sgrad_img}.

%% emission measure is 0.5*normalization/area; area = 15 sq arcsec
%% note 1'' = 0.039 pc, so area = 0.022 pc

\section{Discussion}

We have searched for X-ray emission from young stars located between 
3--23 pc in projection from \sgrastar, and we have detected 2 of 5 known 
isolated, young stars \citep{cot99}, and identified one new example. 
The stars that we detected in our \chandra\ image are exceptional. 
The star associated with the SgrA-A HII region is the only Wolf-Rayet star 
(spectral type WN6) in the sample of \citet{cot99}.
Nitrogen-rich Wolf-Rayet stars (WN) produce very fast winds 
($v\ga1000$ km s$^{-1}$, which produce strong X-ray
emitting shocks. About 75\% of WN stars have 
$L_{\rm X} > 5\times10^{31}$~\ergs\  \citep{wes96}, so the detection of 
the SgrA-A WN6 star is not surprising (Table~\ref{tab:spec}). 

The two most luminous X-ray sources, \alvin\ and H2, are the brightest 
infrared sources in Table~\ref{tab:prop}.  We compared their spectra
(Fig.~\ref{fig:spex}) to those of the Pistol star
\citep{fig98} and to the atlases of \citet{mor96} and Hanson, 
Conti, \& Rieke (1996).\nocite{han96}. 
Our low-resolution spectra cannot be used to classify stars with the
detail achievable with high-resultion optical spectra. However, the line 
emission from Br-$\gamma$, He I, and Mg II in Figure~\ref{fig:spex}
resembles that from the
Pistol star and other B[e] supergiants, Of stars, and LBVs. All of these
spectral types represent massive stars that have evolved off of the main 
sequence, but not yet reached the Wolf-Rayet phase, so the similarities 
are understandable. For H2, our results match the conclusions of 
\citet{fig95} and \citet{cot99}. 

Our suggestion that these stars are B[e], Of, or LBV stars suggests that they
are unusually
luminous. To confirm this, we estimate the intrinsic colors and luminosities 
of \alvin\ and H2 as follows. 
First, we assume that they lie near the Galactic center ($D$$=$8 kpc), because 
(1) the interstellar absorption columns that we measure from their X-ray 
spectra (Table~\ref{tab:spec}) are quite similar to the value toward 
\sgrastar, $N_{\rm H} = 6\times10^{22}$ cm$^{-2}$, and (2) if we ignore 
the heavily-absorbed half of the Galaxy more distant 
than 10~kpc, $\sim80$\% of the Galactic stellar mass 
along this line-of-sight is located within $\pm$200 pc of the Galactic center 
\citep{lzm02}. The distance modulus is therefore 14.5.
Second, we estimate the reddening using the relationship $N_{\rm H}/A_V = 
1.79 \times 10^{21}~{\rm cm}^{-2}$ from \citet{ps95}, and the extinction 
law from from \citet{mon01}. For \alvin, taking 
$N_{\rm H} = (4.4\pm0.3)\times10^{22}$~cm$^{-2}$, we find 
$A_H = 4.6\pm0.3$, and $A_K = 2.9 \pm 0.2$. Its intrinsic color is 
$(H - K_s)_0 = -0.4 \pm 0.1$, and absolute $M_K = -9.5\pm0.2$. For H2, 
$N_{\rm H} = (7.7\pm1.4)\times10^{22}$~cm$^{-2}$, we find 
$A_H = 7.5\pm1.4$, and $A_K = 4.8 \pm 0.9$. Its intrinsic color is 
$(H - K_s)_0 = -0.6 \pm 0.5$, and $M_K = -10.1\pm0.9$. In both cases,
the colors are within 1$\sigma$ of $(H-K)_0 = -0.3 \pm 0.1$ expected for 
O supergiant stars \citep{weg94}. However, the colors are inconsistent with 
the $(H-K)_0 \ga 0.8$ infrared excesses seen from B[e] stars in the 
Magellanic clouds \citep{zic86}. Therefore, we tentatively classify 
\alvin\ and H2 as either Of or candidate LBV stars. 

We estimate the bolometric luminosities of these stars
by assuming that they have temperatures comparable to those
of the Pistol star and other hot, massive stars in transition, 
$T \ga 1.5\times10^{4}$ K \citep{mor96,fig98,cla03}. 
The bolometric corrections to the absolute $K_s$ magnitudes will then be 
given by 
$BC_K = -7.1\log T + 28.8$ \citep{blu95}, so that $BC_K \la -0.9$.
Therefore, we find upper limits on the bolometric magnitudes 
for \alvin\ of $M_{\rm Bol} \la -10.2$, and for H2 of 
$M_{\rm Bol} \la 10.1$. We converted these to {\it lower}
limits on the luminosities (listed in Table~\ref{tab:lxlbol}), and find for 
\alvin\ that $L_{\rm Bol} > 10^{6.1} L_\odot$, and for H2 that
$L_{\rm Bol} > 10^{5.9} L_\odot$. Their true luminosities
may be significantly higher. For example, assuming $T$=30,000 K, as 
would be appropriate for an Of star, the bolometric luminosity would be 
a factor of 7 higher.

\subsection{Radio, Mid-Infrared, and X-ray Emission}

Although \alvin\ and H2 are very luminous stars, it is nonetheless
surprising that they are so bright in the radio, mid-infrared, and 
X-rays. Free-free emission in the winds of massive, evolving stars stars 
should make them mJy radio sources from the distance to the Galactic 
center \citep[e.g.,][]{scu98,dw02}. 
The point-like radio source consistent with \alvin\ is consistent with 
a stellar wind. We can estimate 
the required mass-loss rate by inverting Eq.~4 in \citet{scu98}:
\begin{displaymath}
\dot{M} = 2\times10^{-7} \left[ S_\nu d_{\rm kpc}^2
\nu_{10~{\rm GHz}}^{-0.6} T_4^{0.1} \right]^{3/4} 
\frac{\mu_e v_{\infty}}{10^2~{\rm km/s}}~M_\odot~{\rm yr}^{-1}.
\end{displaymath}
If we assume the electron temperature in units of $10^4$~K is $T_4 = 1$, 
the mean atomic weight 
per free electron is $\mu_e = 1.3$, and the terminal velocity of the wind
is $v_\infty = 200$~km s$^{-1}$ as appropriate for an LBV 
\citep[e.g., P Cyg in][]{scu98}, then $S_{8.4~{\rm GHz}} = 2.3$ mJy
implies $\dot{M} = 2\times10^{-5}$~\msun\ yr$^{-1}$. This is similar
to the mass loss rates observed for LBVs \citep[see also, e.g.,][]{cla03}.
A higher velocity of $v_{\infty} \approx 1000$ km s$^{-1}$, 
as appropriate for an Of star, would imply a mass loss rate 
$>10^{-4}$~\msun\ yr$^{-1}$.
This rate is much higher than is typically seen from Of stars \citep{scu98}.
However, that we have not measured the radio spectral index
of \alvin, so it is possible that most of the flux is non-thermal, 
the above mass-loss rates are upper limits. To determine the nature of 
the radio emission from \alvin, either the radio spectral index must 
be measured, or the mass-loss rate must be obtained from a 
high-resolution infrared spectrum.

The radio flux of the extended nebula
around H2 is orders of magnitude too bright to be produced by its stellar 
wind. \citet{zhao93} model it as an unusually dense ($n_e \sim 10^3$ cm$^{-3}$)
HII region illuminated by a Lyman continuum flux equivalent to that of an
O7 zero-age main sequence star. 

The detection of both stars 
in the MSX mid-infrared survey demonstrates that they are surrounded 
by warm gas and dust. \citet{egan02} and \citet{cla03} have identified 
dusty ring nebulae around several known and candidate LBV stars, and 
hypothesize that they were produced by past episodes of rapid mass loss. 
Surprisingly, the mid-infrared counterparts to our Galactic center stars are 
$\ga$10 and 100 times more luminous at 14.7 and 21.3 $\mu$m, respectively, 
than the nebulae around the other candidate LBVs. 
Therefore, we suspect either that the 
mid-infrared flux near \alvin\ and H2 represents interstellar dust, 
or that it is the unresolved emission from several massive, dusty stars.
Observations with the {\it Spitzer Space Telescope} should resolve
this issue.

The detection of \alvin\ and H2 with \chandra\ is particularly intriguing 
because Of and LBV stars are not always 
bright X-ray sources. In Table~\ref{tab:lxlbol}, we tabulate the values of 
$L_{\rm X}/L_{\rm Bol}$ for several of the massive stars listed 
in \citet{mor96} and \citet{fig99}. 
%,\footnote{Only a few supergiant 
%B[e] stars are known in our Galaxy \citep{lam98}, and sensitive X-ray 
%measurements of these stars are not available.}
The measurements of $L_{\rm X}$ were taken with instruments 
sensitive to different energy ranges, so some caution should be used in 
comparing the results, as described in the notes to the table. 
In general, we have extrapolated the fluxes using conservative assumptions 
that serve to lessen 
the dispersion in $L_{\rm X}/L_{\rm Bol}$. Therefore, we expect the 
range in these values is not caused by selection effects, but represents 
intrinsic differences in the efficiency of X-ray production in these sources.
We find that there is at least a factor of 3000 range in the values, and
that the scatter is not correlated with spectral type. The LBV $\eta$ Car and 
the O6 f star HD 108 lie near the X-ray-bright end of the distribution
with $L_{\rm X}/L_{\rm Bol} = -5.3$ and $-6.2$, respectively. The LBV P 
Cyg and the O4.5 If star HD 190429 lie at the faint end with 
$L_{\rm X}/L_{\rm Bol} < -8.5$.

\begin{deluxetable}{lccccc}
\tabletypesize{\scriptsize}
\tablecolumns{6}
\tablewidth{0pc}
\tablecaption{X-ray and Bolometric Luminosities of Transitional Massive Stars\label{tab:lxlbol}}
\tablehead{
\colhead{Star} & \colhead{Type} & \colhead{$L_{\rm Bol}$} & 
\colhead{$L_{\rm X}$} &  \colhead{$L_{\rm X}/L_{\rm Bol}$} & \colhead{Ref.} \\
\colhead{} & \colhead{} & \multicolumn{2}{c}{(log$[$ erg s$^{-1}$$]$)} & 
\colhead{(log)} & \colhead{}
}
\startdata
\alvin\ & & 39.7 & 33.3 & --6.4 & 1 \\
H2 & & 39.5 & 33.1 & --6.4 & 1 \\
%H8 & & 38.9 & $<$32 & $<$ --6.9 & 1 \\
%H5 & & 38.6 & $<$32 & $<$ --6.6 & 1 \\
%SgrA-A & WN 6 & 38.3 & 31.7 & --6.6 & 1 \\
%SgrA-D & & 39.4 & $<$32 & $<$ --7.4 & 1 \\ [5pt]
P Cyg & LBV & 39.5 & $<$31.0\tablenotemark{a} & $<$ --8.5 & 2,3 \\
$\eta$ Car & LBV & 40.1 & 34.8 & --5.3 & 2,4 \\
Pistol & LBVc & 40.2 & $<$32.0\tablenotemark{b} & $<$ --8.2 & 1,5 \\
HD 108 & O6 f & 39.2 & 33.0 & --6.2 & 6 \\
HD 190429 & 04.5 If$^+$ & 39.7 & 31.1\tablenotemark{a,c} & --8.6 & 7,8,9 \\ %RASS; d=1.7 pc from \citep{scu98}
HD 152408 & O8:Iafpe & 39.4 & $<$31.7\tablenotemark{a} & $<$ --7.7 & 7,10 \\ 
HD 151804 & O8Iaf & 39.7 & 31.9\tablenotemark{a,c} & --7.8 & 7,11,12  \\ % 0.02 c/s ROSAT PSPC catalog, d=1.9 kpc from Lamers and Leitherer
\enddata
\tablecomments{The bolometric luminosities for the two sources in this work
are estimated by assuming values for the extinction and for the bolometric 
correction for each star, as described in the text. The X-ray luminosties are 
extrapolated from the observed values into the 0.5-8.0 keV band, except
where indicated. The 0.5--8.0 keV luminosities of massive stars are usually
dominated by $kT \approx 0.5$ keV plasma that would be undetectable through
the absorption toward the Galactic center; at most 10\% of $L_{\rm X}$ 
should be produced by hotter 1~keV plamsa. Therefore, both luminosity values 
are uncertain by roughly an order of magnitude.}
\tablenotetext{a}{The luminosities are reported in the \rosat\ 0.1-2.0 keV 
band for these sources. For direct comparison with the other sources, one 
would want to use the 0.5--8.0~keV luminosity, which will in general be 
a factor of $\sim$10 smaller.}
\tablenotetext{b}{We derived an upper limit to the X-ray luminosity for 
the Pistol star using archival \chandra\ data, in the same manner as we 
did for the other stars in our survey.}
\tablenotetext{c}{We assumed that 0.01 count s$^{-1}$ in the \rosat\ PSPC 
equals $1\times10^{-13}$~\ergcms, which is appropriate for a 0.5 keV thermal
plasma spectrum absorbed by $5\times10^{20}$ cm$^{-2}$ of H
\citep[see, e.g.,][]{ber96}. The true correction can vary by a factor of 4 for 
reasonable choices of the assumed absorption and temperature.}
\tablerefs{(1) This work; (2) \citet{hd94}; (3) \citet{bw00}; 
(4) \citet{sew01}; (5) \citet{fig99}; (6) \citet{naz04}; (7) \citet{mor96};
(8) \citet{scu98}; (9) \citet{vog99}; (10) \citet{ber96}; 
(11) \rosat\ PSPC Pointed Observations catalog; (12) \citet{ll93}.}
\end{deluxetable}

\alvin\ and H2 lie at the high end of the scatter in $L_{\rm X}/L_{\rm Bol}$
in Table~\ref{tab:lxlbol}, which raises the question of why they are so
bright in X-rays. There are several possible explanations. 
One is that the X-ray emitting shocks form at 
varying optical depths within the winds of massive stars, so that 
in some cases the X-ray emission is absorbed before it can reach the 
observer. This has been proposed to explain why Carbon-type
WR stars are faint X-ray sources, whereas Nitrogen-type WR stars are 
bright in X-rays \citep{osk03}. However, changes in the optical 
depth of the wind from \alvin\ are unlikely to explain the 
X-ray flare observed in 2004 July (\S2.1). Alternatively, the X-rays 
from \alvin\ and H2 could be produced when their winds collide with those
of fainter companions. For instance, it has been proposed that the high X-ray 
luminosity of $\eta$ Car is produced because it is a binary, and one of the
components has a fast, $v\sim 1000$ km s$^{-1}$ wind that collides with 
the slow, dense wind of its companion \citep{sew01}.
Finally, these stars could have black hole or neutron star companions 
that produce X-rays as they accrete from the 
wind of the supergiant. This would make \alvin\ and H2 analogous to faint
high-mass X-ray binaries such as A 0535+26 and X Per, which have 
similar X-ray spectra and luminosities \citep{ds98,orl04},
and exhibit prominent He I and Br-$\gamma$ emission in their spectra
\citep{han96}.
Both colliding-wind binaries and accreting compact objects exhibit 
X-ray flares similar to that seen from \alvin.
Further high-resolution spectroscopic observations will allow us to model 
the winds from \alvin\ and H2, and to search for evidence for binary 
companions.

Given the scatter in $L_{\rm X}/L_{\rm Bol}$ in Table~\ref{tab:lxlbol}, 
it is not surprising that we did not detect X-ray emission from the 
remaining three stars from \citet{cot99}. The spectra from
\citet{cot99} suggest that these stars are all in similar evolutionary
states as \alvin\ and H2. If we make similar assumptions 
to compute their $L_{\rm Bol}$, we find that the remaining stars are a 
only factor of a few less luminous than \alvin\ and H2. However, if we 
assume that they are producing X-rays with characteristic temperatures of 
$\la$0.5 keV, then most of the X-ray emission would be undetectable 
through the absorption toward the Galactic center, and from the upper 
limits in \S2.1 we find $\log(L_{\rm X}/L_{\rm Bol}) < -6$. 
These limits are not particularly interesting in the context of 
X-ray emission from massive stars. Nonetheless, we can confidently state 
these stars are at least an order of magnitude fainter than 
\alvin\ and H2 in the observable 2--8 keV band (Table~\ref{tab:prop}), 
which highlights the oddity of the latter two sources.

\subsection{Implications for Star Formation}

Hot, post-main sequence Wolf-Rayet, B[e], Of, 
and LBV stars similar to those in Table~\ref{tab:prop} are thought to have 
initial masses of $M \ga 35$ \msun\ \citep{zic86,fig98}, 
and ages of only a few million years. As a result, such stars are excellent
tracers of recent star formation. 

Surprisingly, unlike the 
Pistol star and the massive stars in the central parsec around 
\sgrastar\ \citep{fig98,pau03}, none of the stars in Table~\ref{tab:prop}
 are associated with any known young star 
cluster. However, clusters less massive than the Arches and Quintuplet would 
be difficult to identify against the dense background of stars in the 
Galactic center 
using previous wide-field infrared surveys like 2MASS \citep{pz02}. 
Therefore, it is important to carry out photometric and narrow-line surveys 
to search for other young stars in the vicinity of H2 and \alvin, to 
establish whether they are in fact isolated. 

If they are isolated, then there are two 
possible explanations. First, it has been suggested that the 
young stars in the central parsec are the remnants of a cluster that
was formed $\sim$10 pc from \sgrastar\ and settled into the Galactic 
center \citep{ger01}, and stars like H2 and \alvin\ could have 
been stripped from this hypothetical cluster. If this is the case,
then hundreds of other young stars still await discovery in the
central 10 parsecs around \sgrastar\ \citep[e.g.,][]{pmg03, km03}.
Alternatively, these massive, young stars may well have formed in 
isolation, as has been 
proposed by \citet{cot99}. This hypothesis
would imply that stars form in associations that have masses ranging from 
100 to $10^{4}$~\msun. Such a conclusion would be
surprising, since the high turbulent velocities of molecular clouds, 
the large tidal forces produced by \sgrastar\ and the surrounding stellar cusp,
and possibly mG magnetic fields concentrated near the Galactic center 
by in-falling material should collude to make the Jeans mass closer
to the upper end of this range \citep{mor93}.

\acknowledgments
We thank D. Figer, A. Cotera, E. Feigelson, L. Townsley, and J. S. Clark
for very helpful discussions about the natures of these young stars,
and S. Eikenberry for helpful advice about identifying counterparts to
the X-ray sources. 
MPM was supported through a Hubble Fellowship grant (program number
HST-HF-01164.01-A) from the Space Telescope Science Institute, which is 
operated by the Association of Universities for Research in Astronomy,
Incorporated, Under NASA contract NAS5-26555.
AJB acknowledges support from NASA through the Spitzer Fellowship Program.
WNB was supported by NSF CAREER Award 9983783.
This publication makes use of data products from the Two Micron All Sky 
Survey, which is a joint project of the University of Massachusetts and 
the Infrared Processing and Analysis Center/California Institute of 
Technology, funded by the NASA and the NSF.

\end{document}